\documentstyle[onecolumn]{mn}

\newif\ifAMStwofonts
\DeclareMathAlphabet\EuFrak{U}{euf}{m}{n}
\SetMathAlphabet\EuFrak{bold}{U}{euf}{b}{n}
  \DeclareFontFamily{U}{euf}{}%
  \DeclareFontShape{U}{euf}{m}{n}{<-6>eufm5<6-8>eufm7<8->eufm10}{}%
  \DeclareFontShape{U}{euf}{b}{n}{<-6>eufb5<6-8>eufb7<8->eufb10}{}%
\newcommand{\mathfrak}{\EuFrak}
\RequirePackage{keyval,graphics}
\RequirePackage{subfigure}
\RequirePackage{graphicx}
\RequirePackage{graphics,psfrag}
\DeclareSymbolFont{AMSa}{U}{msa}{m}{n}
\DeclareSymbolFont{AMSb}{U}{msb}{m}{n}
\DeclareSymbolFontAlphabet{\mathbb}{AMSb}

\ifoldfss
  \ifCUPmtlplainloaded \else
    \NewTextAlphabet{textbfit} {cmbxti10} {}
    \NewTextAlphabet{textbfss} {cmssbx10} {}
    \NewMathAlphabet{mathbfit} {cmbxti10} {} % for math mode
    \NewMathAlphabet{mathbfss} {cmssbx10} {} %  "   "    "
  \fi
  \ifAMStwofonts
    \ifCUPmtlplainloaded \else
      \NewSymbolFont{upmath} {eurm10}
      \NewSymbolFont{AMSa} {msam10}
      \NewMathSymbol{\upi}     {0}{upmath}{19}
      \NewMathSymbol{\umu}     {0}{upmath}{16}
      \NewMathSymbol{\upartial}{0}{upmath}{40}
      \NewMathSymbol{\leqslant}{3}{AMSa}{36}
      \NewMathSymbol{\geqslant}{3}{AMSa}{3E}

      \let\leq=\leqslant 
      \let\geq=\geqslant 
    \fi
  \fi
\fi % End of OFSS

\ifnfssone
  \newmathalphabet{\mathit}
  \addtoversion{normal}{\mathit}{cmr}{m}{it}
  \addtoversion{bold}{\mathit}{cmr}{bx}{it}
  \newmathalphabet{\mathbfit} % math mode version of \textbfit{..}
  \addtoversion{normal}{\mathbfit}{cmr}{bx}{it}
  \addtoversion{bold}{\mathbfit}{cmr}{bx}{it}
  \newmathalphabet{\mathbfss} % math mode version of \textbfss{..}
  \addtoversion{normal}{\mathbfss}{cmss}{bx}{n}
  \addtoversion{bold}{\mathbfss}{cmss}{bx}{n}
  \ifAMStwofonts
    \ifCUPmtlplainloaded \else
      \UseAMStwoboldmath
      \makeatletter
      \new@mathgroup\upmath@group
      \define@mathgroup\mv@normal\upmath@group{eur}{m}{n}
      \define@mathgroup\mv@bold\upmath@group{eur}{b}{n}
      \edef\UPM{\hexnumber\upmath@group}
      \new@mathgroup\amsa@group
      \define@mathgroup\mv@normal\amsa@group{msa}{m}{n}
      \define@mathgroup\mv@bold\amsa@group{msa}{m}{n}
      \edef\AMSa{\hexnumber\amsa@group}
      \makeatother
      \mathchardef\upi="0\UPM19
      \mathchardef\umu="0\UPM16
      \mathchardef\upartial="0\UPM40
      \mathchardef\leqslant="3\AMSa36
      \mathchardef\geqslant="3\AMSa3E

      \let\leq=\leqslant 
      \let\geq=\geqslant 
    \fi
  \fi
\fi % End of NFSS release 1

\ifnfsstwo
  \DeclareMathAlphabet{\mathbfit}{OT1}{cmr}{bx}{it}
  \SetMathAlphabet\mathbfit{bold}{OT1}{cmr}{bx}{it}
  \DeclareMathAlphabet{\mathbfss}{OT1}{cmss}{bx}{n}
  \SetMathAlphabet\mathbfss{bold}{OT1}{cmss}{bx}{n}
  \ifAMStwofonts
    \ifCUPmtlplainloaded \else
      \DeclareSymbolFont{UPM}{U}{eur}{m}{n}
      \SetSymbolFont{UPM}{bold}{U}{eur}{b}{n}
      \DeclareSymbolFont{AMSa}{U}{msa}{m}{n}
      \DeclareMathSymbol{\upi}{0}{UPM}{"19}
      \DeclareMathSymbol{\umu}{0}{UPM}{"16}
      \DeclareMathSymbol{\upartial}{0}{UPM}{"40}
      \DeclareMathSymbol{\leqslant}{3}{AMSa}{"36}
      \DeclareMathSymbol{\geqslant}{3}{AMSa}{"3E}

      \let\leq=\leqslant 
      \let\geq=\geqslant 
    \fi
  \fi
\fi % End of NFSS release 2

\ifCUPmtlplainloaded \else
  \ifAMStwofonts \else % If no AMS fonts
    \def\upi{\pi}
    \def\umu{\mu}
    \def\upartial{\partial}
  \fi
\fi

\title[Two-integral DFs for axisymmetric systems]
  {Two-integral distribution functions for axisymmetric systems}
\author[Z.~Jiang and L.~Ossipkov]
  {Zhenglu Jiang$^1$\thanks{mcsjzl@mail.sysu.edu.cn} 
       and Leonid Ossipkov$^2$\thanks{leo@dyna.astro.spbu.ru}
\\
  $^1$Department of Mathematics, Zhongshan University, 
              Guangzhou 510275, China\\
  $^2$Saint~Petersburg State University, Staryj Peterhof, Saint~Petersburg 198504, Russia}
\date{Accepted 2007 May 16. Received 2007 May 15; in original form 2007 April 18}
\pagerange{\pageref{firstpage}--\pageref{000}}
\volume{379}
\pubyear{2007}

\begin{document}

\label{firstpage}

\maketitle

\begin{abstract}
 Some formulae are presented for finding two-integral distribution functions (DFs) 
which depends only on the two classical integrals of the energy and the 
magnitude of the angular momentum with respect to the axis of symmetry 
for stellar systems with known axisymmetric densities. 
They come from an combination of the ideas of 
 Eddington and Fricke and they 
are also an extension of those shown by Jiang and Ossipkov for  
finding anisotropic DFs for spherical galaxies. 
The density of the system is required to be expressed as  
  a sum of products of functions of the potential and 
  of the radial coordinate. 
The solution corresponding to this type of density  
is in turn a sum of products of functions of the energy and of the 
magnitude of the angular momentum about the axis of symmetry. 
The product of the density and its radial velocity dispersion   
can be also expressed as  
  a sum of products of functions of the potential and 
 of the radial coordinate.  It can be further known that the density 
multipied by its rotational velocity dispersion  is 
equal to a sum of products of functions of the potential and 
 of the radial coordinate 
minus the product of the density and the square of its mean rotational velocity.
These formulae can be applied to the Binney and the Lynden-Bell models. 
An infinity of the odd DFs for the Binney model can be also found under the assumption of 
the laws of the rotational velocity.
\end{abstract}


\section{Introduction}
\label{intro}
It is the most straightforward to  
construct self-consistent stellar systems by means of finding 
distribution functions for a stellar system with a known gravitational potential.   
Once the potential of the system is known,  
the mass density $\rho$ of the system can be uniquely detemined via Poisson's 
equation generated by the well-known Newtonian gravitational law, and  
the structure of the stellar orbits can be also obtained according to Newton's 
equations of motion. The system is hence constructed from 
the structure of the orbits in the potential. This construction  
is also the so-called ``from $\rho$ to $f$'' approach for finding a 
self-consistent distribution function $f$ 
(Binney \& Tremaine 1987, hereafter BT; Hunter \& Qian 1993). 
Since the mass density is the integration 
of the distribution function (hereafter DF) over the velocity variable 
in the phase space of the system, 
 the problem of finding the DF is that of solving an 
integral equation. 
In a system with a known spherical potential, there is a unique isotropic DF shown by Eddington (1916) 
and many different anisotropic DFs obtained by many other outstanding astronomers   
(e.g., Camm 1952; Bouvier 1962, 1963; Kuzmin \& Veltmann 1967, 1973;  
Veltmann 1961, 1965, 1979, 1981;  Kent \&Gunn 1982; Dejonghe 1986, 1987; 
Dejonghe \& Merritt 1988). 
Recently, a method was presented by Jiang \& Ossipkov (2007) for  
finding anisotropic distribution functions for spherical galaxies. 
This is an combination of Eddington's (1916) formula 
and Fricke's (1952) expansion idea.  Of course, 
they can be also regarded as simply
an extension of the idea of Eddington. 

Fricke's (1952) expansion idea is that  
DFs which are products of the two powers 
of the energy and the square of the angular momentum about the axis of symmetry 
correspond to densities which are proportional to products of the potential 
and the radial coordinate for axisymmetric systems.  
Hence the DF for the system can be obtained 
by first expressing the density as a function of 
the potential and the radial coordinate, 
and then expanding as a power series. 
According to the maximum entropy principle, 
there are an infinity of the most probable two-integral DFs  
for a given mass distribution (see Dejonghe 1986). 
Thus there may be an infinity of two-integral DFs corresponding to 
any given mass density  in axisymmetric stellar systems.  

In the literature, there are a number of different axisymmetric models for galaxies 
(e.g., Miyamoto 1971; Bagin 1972; Miyamoto \& Nagai 1975; Nagai \& Miyamoto 1976; Kutuzov \& Ossipkov 1980, 1986, 1988; 
Evans 1993, 1994; Kutuzov 1995; Jiang 2000; Jiang \& Moss 2002; Jiang et.~al 2002; Jiang \& Ossipkov 2006; 
Ossipkov \& Jiang 2007) 
and different integral transformation techniques are used to 
obtain the solution of the problem of findng the two-integral 
DFs of axisymmetric stellar systems
(e.g. Lynden-Bell 1962; Hunter 1975; Kalnajs 1976; Dejonghe 1986) but 
there is a fatal obstacle of requiring not only 
the validity of these transformations of the density 
but also the complex analyticity 
of a density-related integral kernel to complex arguments. 
It is a great progress that 
the contour integral of Hunter \& Qian (1993) is 
used to find the two-integral DFs for axisymmetric systems.  
However,  this is still a complex integral suitable for analytic densities and its    
singularities are required to satisfy some conditions.   

Therefore the above ideas of Eddington and Fricke are still very useful to the problem of finding the DFs. 
The main aim of this paper is to present some formulae for finding two-integral DFs 
in axisymmetric systems by an combination of  both the  
 Eddington formula and the Fricke expansion (see above).
The fundamental integral equations of the problem of finding the two-integral DFs are given   
in Section \ref{finteq} and some new formulae 
of the two-integral DFs for stellar systems
with known axisymmetric densities are shown 
in Section  \ref{dfas}.
These DFs depend only on the two integrals of the energy and the 
 magnitude of the angular momentum about the symmetry axis.   
These formulae are also an extension of those shown by Jiang \& Ossipkov (2007) for  
finding anisotropic distribution functions for spherical galaxies. 
A type of two-integral DF which is  
a sum of products of functions only of the energy  
and powers of the magnitude of the angular momentum with respect to the axis of symmetry 
is derived in Section \ref{edf} and  
another, which is a sum of products 
of functions only of a special variable  
and powers of the magnitude of the angular momentum about the axis of symmetry,
in Section \ref{edfq}. 
More general formulae are given in the last part of Section \ref{dfas}.  
Various formulae of the velocity dispersions for such models of these DFs 
are also  shown in all the three parts of Section \ref{dfas}. 
For the gravitational potentials having no upper bound and tending to $+\infty,$ 
the similar formulae of the DFs are also shown in Section \ref{mgpinf}.
Their application is given in Section \ref{aac}. 
Section {\ref{con} is a summary and conclusion.

\section{The fundamental integral equations}
\label{finteq}
Assume that $\Phi$ and $E$ are, respectively, the gravitational potential 
and the energy of a star in a stellar system. 
As in BT, one can choose a  constant $\Phi_0$ 
such that the system has only stars of the energy $E< \Phi_0,$  
and then define $\psi=-\Phi+\Phi_0$ and $\varepsilon=-E+\Phi_0.$ In this system, 
the two physical quantities $\psi$ and $\varepsilon$ are usually 
 called the relative potential and energy, respectively. 
Obviously, $\varepsilon=0$ is a relative energy of escape from the system. 
Given a stellar system with the relative potential $\psi=\psi({\bf r}),$ 
its mass density $\rho=\rho({\bf r})$ can be obtained by using Poisson's equation 
and its distribution function $f=f({\bf r},{\bf v})$ satisfies the following integral equation  
\begin{equation}
-\nabla^2\psi =4\pi G \rho=4\pi G\int fd^3{\bf v}, \label{int0}
\end{equation}
where ${\bf r}$ is a position vector, ${\bf v}$ is a velocity vector, 
$G$ is the gravitational constant. The cylindrical 
polar coordinates $(R,\varphi,z)$ are generally used with  
the $z$-axis being that of symmetry for an axisymmetric system. 
The velocity in cylindrical 
polar coordinates $(R,\varphi,z)$ is usually denoted by 
${\bf v}=(v_R,v_\phi,v_z)$ and $L_z$ is the component of angular 
momentum about the $z$-axis.  Then $L_z=R v_\phi$ and 
it is well known that the relative energy $\varepsilon$ 
and the $z$-axis angular momentum $L_z$ are two isolating integrals for any orbit in 
the axisymmetric system. Hence, by the Jeans theorem, the DF of a steady-state stellar system 
in an axisymmetric potential can be expressed as a non-negative function of $\varepsilon$ 
and $L_z,$  denoted by $f(\varepsilon,L_z),$ and then 
for an axisymmetric system,  (\ref{int0}) can be rewritten as 
\begin{equation}
 -\frac{1}{R}\frac{\partial}{\partial R}\left(R\frac{\partial\psi}{\partial R}\right)
-\frac{\partial^2\psi}{\partial z^2}=4\pi G\rho=4\pi G\int  f(\varepsilon,L_z)d^3{\bf v}. \label{inta1}
\end{equation}
Let $f_+(\varepsilon,L_z)=[f(\varepsilon,L_z)+f(\varepsilon,-L_z)]/2.$ Then, 
by $\varepsilon=\psi-(1/2)(v_R^2+v_\phi^2+v_z^2),$ 
the integral given by (\ref{inta1}) can be expressed as 
\begin{equation}
 \rho=\frac{4\pi}{R}\int_0^\psi\left[\int_0^{R\sqrt{2(\psi-\varepsilon)}} 
f_+(\varepsilon,L_z)dL_z\right]d\varepsilon \label{inta2}
\end{equation}
since the system has only stars with $\varepsilon>0,$ that is, $f(\varepsilon,L_z)=0$ for $\varepsilon>0.$ 
This implies that a given density determines $f_+(\varepsilon,L_z)$ which is 
just the part of the DF that is even in $L_z.$ 
Hence $f_+(\varepsilon,L_z)$ is usually called  the even DF.

Once $f_+(\varepsilon,L_z)$ is known, $f(\varepsilon,L_z)$ can be obtained 
under some suitable assumptions such as the isotropy of the two-integral DFs (see BT) and 
the maximum entropy of the most probable two-integral DFs  
(see Dejonghe 1986), and further $\rho$ can be easily calculated 
by integration and $\psi$ by solving Poisson's equation for the axisymmetric system. 
 The inverse problem that  is now investigated is 
how to derive the two-integal DF $f_+(\varepsilon,L_z)$ from the density $\rho$ for any axisymmetric system.  
Different classes of the two-integal DFs will be below shown,  
which are derived from axisymmetric density profiles for galaxies,  
by combining some functions only of $\varepsilon$ 
(or $Q\equiv\varepsilon-L_z^2/(2R_a^2)$) 
with some functions of the form $L_z^{2n\beta_n}$ 
where $R_a$ is a scaling radius, $n$ is an integer greater than $-2$ 
and $\beta_n$ is a constant such that $n\beta_n>-1.$ 

\section{Two-integral DFs for Axisymmetric Systems}
\label{dfas}
In this section various formulae for the even DFs of axisymmetric systems are obtained 
from axisymmetric density profiles of different forms 
and their radial and rotational velocity dispersions are expressed 
in a simple form. 

\subsection{DFs of the form $\sum\limits_{n=0}^{m}L_z^{2n}h_n(\varepsilon)$}
\label{edf}
Note that the integral on the right side of (\ref{inta2}) is in fact a function of the relative potential $\psi$ 
and the radial coordinate $R.$ 
Hence, by (\ref{inta2}), the mass density $\rho$ can be regarded as a function 
depending on the relative potential $\psi$ and the radial coordinate $R.$ 
Let  $\rho=\rho({\bf r})$ be below denoted by $\rho(\psi,R).$ 
Assume that $f_+(\varepsilon,L_z)=\sum\limits_{n=0}^{m}L_z^{2n}h_n(\varepsilon)$ 
and that $\rho(\psi,R)=\sum\limits_{n=0}^{m}R^{2n}\tilde{\rho}_n(\psi).$ Then, by (\ref{inta2}), 
it is easy to see that 
\begin{equation}
 \sum\limits_{n=0}^{m}R^{2n}\tilde{\rho}_n(\psi)
=\sum\limits_{n=0}^{m}\frac{4\pi 2^{n+1/2}R^{2n}}{2n+1}
\int_0^\psi h_n(\varepsilon) (\psi-\varepsilon)^{n+1/2}d\varepsilon. \label{inta3}
\end{equation}
It follows from (\ref{inta3}) that 
\begin{equation}
\tilde{\rho}_n(\psi)=\frac{4\pi 2^{n+1/2}}{2n+1}
\int_0^\psi h_n(\varepsilon) (\psi-\varepsilon)^{n+1/2}d\varepsilon. \label{inta4}
\end{equation}
Assume that $(d^j\tilde{\rho}_n(\psi)/d\psi^j)_{\psi=0}=0$  for all 
$j\in \{0,1,\dots,n\}$ and all $n\in \{0,1,2,\dots,m\}.$
By taking the $(n+1)$st derivative of (\ref{inta4}) and using Abel's integral equation, 
one can get 
\begin{equation}
 h_n(\varepsilon)=\frac{1}{(2\pi)^{3/2}2^n\Gamma(n+\frac{1}{2})}\frac{d}{d\varepsilon}\int_0^\varepsilon
\frac{d^{n+1}\tilde{\rho}_n(\psi)}{d\psi^{n+1}}\frac{d\psi}{\sqrt{\varepsilon-\psi}}
 \label{solutiona}
\end{equation}
for $n=0,1,2,\dots,m.$   
Hence it can be easily known from (\ref{solutiona}) that 
\begin{equation}
f_+(\varepsilon,L_z)=\frac{1}{(2\pi)^{3/2}}
\sum\limits_{n=0}^{m}\frac{L_z^{2n}}{2^n\Gamma(n+\frac{1}{2})}\frac{d}{d\varepsilon}\int_0^\varepsilon
\frac{d^{n+1}\tilde{\rho}_n(\psi)}{d\psi^{n+1}}\frac{d\psi}{\sqrt{\varepsilon-\psi}}
\label{dfa}
\end{equation}
for $\varepsilon>0,$ 
corresponding to the axisymmetric density 
of the form $\rho(\psi,R)=\sum\limits_{n=0}^{m}\tilde{\rho}_n(\psi)R^{2n}.$ 
Also, (\ref{dfa}) can be equivalently rewritten as 
\begin{equation}
f_+(\varepsilon,L_z)=\frac{1}{(2\pi)^{3/2}}
\sum\limits_{n=0}^{m}\frac{L_z^{2n}}{2^n\Gamma(n+\frac{1}{2})}\left[\int_0^\varepsilon
\frac{d^{n+2}\tilde{\rho}_n(\psi)}{d\psi^{n+2}}\frac{d\psi}{\sqrt{\varepsilon-\psi}}
+\frac{1}{\sqrt{\varepsilon}}\left(\frac{d^{n+1}\tilde{\rho}_n(\psi)}{d\psi^{n+1}}\right)_{\psi=0}\right].
\label{dfa-2}
\end{equation}
Furthermore, if it is assumed that $(d^j\tilde{\rho}_n(\psi)/d\psi^j)_{\psi=0}=0$ for all 
$j\in \{0,1,\dots,n+1\}$ and all $n\in \{0,1,2,\dots,m\},$ then, for $\varepsilon>0,$ (\ref{dfa}) can be expressed as 
\begin{equation}
f_+(\varepsilon,L_z)=\frac{1}{(2\pi)^{3/2}}
\sum\limits_{n=0}^{m}\frac{L_z^{2n}}{2^n\Gamma(n+\frac{1}{2})}\int_0^\varepsilon
\frac{d^{n+2}\tilde{\rho}_n(\psi)}{d\psi^{n+2}}\frac{d\psi}{\sqrt{\varepsilon-\psi}}.
\label{dfa0}
\end{equation}

It is worth mentioning that Evans (1994) used Abel transforms to get a similar formula 
of the above even DF in the case for $m=1.$  
His formula of the even DF has just two previous terms 
and is formally expressed as 
\begin{equation}
f_+(\varepsilon,L_z)=\frac{1}{\sqrt{8}\pi^2}\frac{d}{d\varepsilon}\int_0^\varepsilon
\frac{d\tilde{\rho}_0(\psi)}{d\psi}\frac{d\psi}{(\varepsilon-\psi)^{1/2}}
-\frac{L_z^{2}}{2\sqrt{8}\pi^2}\frac{d}{d\varepsilon}\int_0^\varepsilon
\frac{d\tilde{\rho}_1(\psi)}{d\psi}\frac{d\psi}{(\varepsilon-\psi)^{3/2}}.
\label{divdf}
\end{equation}
The real integral in the second term on the right side of (\ref{divdf}) is divergent for the stellar systems 
such as the Lynden-Bell model considered below.   

 By (\ref{dfa}), the velocity dispersions $\sigma_R^2(\psi,R)$ and $\sigma_\varphi^2(\psi,R)$ 
can be also found as follows:  
\begin{equation}
\sigma_R^2(\psi,R)=\frac{1}{\rho(\psi,R)}\sum\limits_{n=0}^{m}R^{2n}\int_0^\psi \tilde{\rho}_n(\psi^\prime)d\psi^\prime
\label{vdra}
\end{equation} 
and 
\begin{equation}
\sigma_\varphi^2(\psi,R)=\frac{1}{\rho(\psi,R)}\sum\limits_{n=0}^{m}(2n+1)R^{2n}\int_0^\psi \tilde{\rho}_n(\psi^\prime)d\psi^\prime 
-\bar{v}_\varphi^2
\label{vdvta}
\end{equation} 
for any DF derived from the axisymmetric density 
of the form $\rho(\psi,R)=\sum\limits_{n=0}^{m}\tilde{\rho}_n(\psi)R^{2n},$ here and below, 
$\bar{v}_\varphi$ represents the mean rotational velocity which can be calculated 
under some suitable assumptions.  It can be also known that 
these dispersions (\ref{vdra}) and (\ref{vdvta}) can be obtained directly
according to Hunter's (1977) formulae as follows:  
\begin{equation}
\sigma_R^2(\psi,R)=\sigma_z^2(\psi,R)=\frac{1}{\rho(\psi,R)}\int_0^\psi \rho(\psi^\prime,R)d\psi^\prime
\label{vdrda}
\end{equation} 
and 
\begin{equation}
\sigma_\varphi^2(\psi,R)=\frac{1}{\rho(\psi,R)}
\int_0^\psi \frac{\partial [R\rho(\psi^\prime,R)]}{\partial R}d\psi^\prime-\bar{v}_\varphi^2. 
\label{vdrda}
\end{equation} 

\subsection{DFs of the form $\sum\limits_{n=0}^{m}L_z^{2n}g_n(Q)$} 
\label{edfq} 
A more general expression for the integral in the right side of (\ref{int0})
 can also be derived. 
Put $Q=\varepsilon-L_z^2/(2R_a^2),$ where $R_a$ is a scaling radius, and 
assume that the DF is dependent on $Q$ and $L_z,$ denoted by  $f(Q,L_z),$ 
and that  the system has only stars with $Q>0,$
or equivalently, $f=0$ for $Q\leq 0.$ 
Obviously, $Q\to \varepsilon$ as $R_a \to \infty.$ 
Then, for an axisymmetric system,  (\ref{int0}) can be expressed as 
\begin{equation}
 -\frac{1}{R}\frac{\partial}{\partial R}\left(R\frac{\partial\psi}{\partial R}\right)
-\frac{\partial^2\psi}{\partial z^2}=4\pi G\rho=4\pi G\int  f(Q,L_z)d^3{\bf v}. \label{intaq1} 
\end{equation}
By changing the variables of the integral in (\ref{intaq1}), it follows that 
\begin{equation}
\rho=4\pi \int_0^{\psi}\left[ \int_0^{\sqrt{2(\psi-Q)/(1+R^2/R_a^2)}}  
f_+(Q,L_z)dv_\phi\right] dQ, \label{intaq2} 
\end{equation} 
where $f_+(Q,L_z)=[f(Q,L_z)+f(Q,-L_z)]/2.$ 
Naturally, $f_+(Q,L_z)$ is the even part of $f(Q,L_z).$
Suppose that  $f_+(Q,L_z)=\sum\limits_{n=0}^{m}L^{2n}g_n(Q),$  
and that the mass density has the following form 
\begin{equation}
\rho(\psi,R)=\sum\limits_{n=0}^{m}\hat{\rho}_n(\psi)R^{2n}/(1+R^2/R_a^2)^{n+1/2}.
\label{rho2}
\end{equation}  
Then (\ref{intaq2}) can be expressed as 
\begin{equation}
 \sum\limits_{n=0}^{m}\frac{R^{2n}\hat{\rho}_n(\psi)}{(1+R^2/R_a^2)^{n+1/2}}
=\sum\limits_{n=0}^{m}\frac{4\pi 2^{n+1/2}R^{2n}}{(2n+1)(1+R^2/R_a^2)^{n+1/2}}
\int_0^\psi g_n(Q) (\psi-Q)^{n+1/2}dQ. \label{intaq3}
\end{equation} 
It is easy to see that equation (\ref{intaq3}) gives 
\begin{equation}
\hat{\rho}_n(\psi)=\frac{4\pi 2^{n+1/2}}{2n+1}
\int_0^\psi g_n(Q) (\psi-Q)^{n+1/2}dQ. \label{intaq4}
\end{equation}
Assume that $(d^j\hat{\rho}_n(\psi)/d\psi^j)_{\psi=0}=0$  for all 
$j\in \{0,1,\dots,n\}$ and all $n\in \{0,1,2,\dots,m\}.$
By taking the $(n+1)$st derivative of (\ref{intaq4}) and using Abel's integral equation, 
one can obtain   
\begin{equation}
 g_n(Q)=\frac{1}{(2\pi)^{3/2}2^n\Gamma(n+\frac{1}{2})}\frac{d}{dQ}\int_0^Q 
\frac{d^{n+1}\hat{\rho}_n(\psi)}{d\psi^{n+1}}\frac{d\psi}{\sqrt{Q-\psi}}
 \label{solutionaq}
\end{equation}
for $n=0,1,2,\dots,m.$   
Hence, corresponding to the axisymmetric density $\rho(\psi,R)$ defined by (\ref{rho2}), 
it can be readily shown from (\ref{solutionaq}) that the DF is obtained as  
\begin{equation}
f_+(Q,L_z)=
\sum\limits_{n=0}^{m}\frac{(2\pi)^{-3/2}L_z^{2n}}{2^n\Gamma(n+\frac{1}{2})}\frac{d}{dQ}\int_0^Q
\frac{d^{n+1}\hat{\rho}_n(\psi)}{d\psi^{n+1}}\frac{d\psi}{\sqrt{Q-\psi}}
\label{dfaq}
\end{equation}
or equivalently, 
\begin{equation}
f_+(Q,L_z)=
\sum\limits_{n=0}^{m}\frac{(2\pi)^{-3/2}L_z^{2n}}{2^n\Gamma(n+\frac{1}{2})}\left[\int_0^Q
\frac{d^{n+2}\hat{\rho}_n(\psi)}{d\psi^{n+2}}\frac{d\psi}{\sqrt{Q-\psi}}
+\frac{1}{\sqrt{Q}}\left(\frac{d^{n+1}\hat{\rho}_n(\psi)}{d\psi^{n+1}}\right)_{\psi=0}\right]
\label{dfaq-2}
\end{equation}
for $Q>0.$ Furthermore, if $(d^j\hat{\rho}_n(\psi)/d\psi^j)_{\psi=0}=0$ for 
$j\in \{0,1,\dots,n+1\}$ and all $n\in \{0,1,2,\dots,m\},$
 then for $Q>0,$ (\ref{dfaq}) can be rewritten as 
\begin{equation}
f_+(Q,L_z)=
\sum\limits_{n=0}^{m}\frac{(2\pi)^{-3/2}L_z^{2n}}{2^n\Gamma(n+\frac{1}{2})}\int_0^Q
\frac{d^{n+2}\hat{\rho}_n(\psi)}{d\psi^{n+2}}\frac{d\psi}{\sqrt{Q-\psi}}.
\label{dfaq0}
\end{equation}

Of course,  (\ref{dfaq}) and (\ref{dfaq0})  coincide with (\ref{dfa}) and (\ref{dfa0}), respectively.   
In other words, (\ref{dfa}) and (\ref{dfa0})  are, respectively, 
limits of (\ref{dfaq}) and (\ref{dfaq0}) when $R_a\to \infty.$ 

Dejonghe (1986) studied the mass densities separable in $\psi$ and $R$ and gave the following 
formula of the even two-integral DFs: 
\begin{equation}
f_+(\varepsilon,L_z)=\frac{\Gamma(p+1)\varepsilon^{p-3/2}}{2^{3/2}\pi}\frac{1}{2\pi{\rm i}}
\int\limits_{\beta_0-{\rm i}\infty}^{\beta_0+{\rm i}\infty}
\frac{{\mathfrak M}_{R\to \beta}\{g\}[L_z^2/(2\varepsilon)]^{-\beta/2}}
{\Gamma(1/2-\beta/2)\beta(p-1/2+\beta/2)}d\beta
\label{ddf}
\end{equation}
for any given mass densities $\rho(\psi,R)=\psi^p g(R),$ where ${\rm i}=\sqrt{-1},$ $p\geq 3/2,$ $\beta_0$ is a suitable constant, 
${\mathfrak M}_{R\to \beta}\{g\}$ represents the Mellin transformation of the function $g(R).$ Let $a$ and $b$ be two constants. 
Assume that $-a-b$ is not a natural number. 
When $g(R)=R^{2a}/(1+R^2)^{a+b},$  (\ref{ddf}) is written by Dejonghe (1986) as 
\begin{equation}
f_+(\varepsilon,L_z)=\frac{\Gamma(p+1)\varepsilon^{p-3/2}}{2^{3/2}\pi\Gamma(a+b)}
{\mathbb H}(a,b,p-\frac{1}{2},\frac{1}{2};\frac{L_z^2}{2\varepsilon})
\label{ddf0}
\end{equation}
where ${\mathbb H}(a,b,c,d;x)$ is defined by 
\begin{equation}
{\mathbb H}(a,b,c,d;x)=\frac{1}{2\pi{\rm i}}\int_{C}
\frac{\Gamma(a+s)\Gamma(b-s)}{\Gamma(c+s)\Gamma(d-s)}x^{-s}ds
\label{hfun}
\end{equation} 
with the contour $C$ such that $-a$ are on its left side 
and $b$ on its right side. 
In the case that $a+d$ and $b+c$ are not negative integers, the complex integral (\ref{hfun}) can be calculated 
and expressed as follows. When $0\leq x<1,$ if $a-c$ is a nonnegative integer, then ${\mathbb H}(a,b,c,d;x)=0,$  or else 
${\mathbb H}(a,b,c,d;x)=x^a  \hbox{}_2F_1(a+b,1+a-c;a+d;x)\Gamma(a+b)/[\Gamma(c-a)\Gamma(a+d)];$ 
when $x>1,$ if $b-d$ is a nonnegative integer, then ${\mathbb H}(a,b,c,d;x)=0,$  or else
${\mathbb H}(a,b,c,d;x)=x^{-b}  \hbox{}_2F_1(a+b,1+b-d;b+c;1/x)\Gamma(a+b)/[\Gamma(d-b)\Gamma(b+c)].$ 
Here, $\hbox{}_2F_1$ is a hyergeometric function. 
In particular, ${\mathbb H}(n,1/2,p-1/2,1/2;x)=x^n\hbox{}_2F_1(n+1/2,n-p+3/2;n+1/2;x)/\Gamma(p-n-1/2)$ for $0\leq x<1,$ 
and ${\mathbb H}(n,1/2,p-1/2,1/2;x)=0$ for $x>1.$ 
Therefore, when $a=n$ and $b=1/2,$ the DF given by (\ref{ddf0}) can be rechanged as 
\begin{eqnarray}
f_+(\varepsilon,L_z) 
=\frac{\Gamma(p+1)\varepsilon^{p-3/2}}{2^{3/2}\pi\Gamma(p-n-1/2)}
\left(\frac{L_z^2}{2\varepsilon}\right)^n 
\hbox{}_2F_1(n+\frac{1}{2},n-p+\frac{3}{2};n+\frac{1}{2};\frac{L_z^2}{2\varepsilon})  \nonumber\\ 
=\frac{\Gamma(p+1)\varepsilon^{p-3/2}}{2^{3/2}\pi\Gamma(p-n-1/2)}
\left(\frac{L_z^2}{2\varepsilon}\right)^n\sum\limits_{k=0}^{+\infty}
\frac{\Gamma(n-p+k+3/2)}{\Gamma(n-p+3/2)k!}\left(\frac{L_z^2}{2\varepsilon}\right)^k  \nonumber \\
=\frac{\pi^{-1}\Gamma(p+1)L_z^{2n}\varepsilon^{p-n-3/2}}{2^{n+3/2}\Gamma(n+\frac{1}{2})\Gamma(p-n-\frac{1}{2})}
\left(1-\frac{L_z^2}{2\varepsilon}\right)^{p-n-3/2}  \hspace*{1.7cm}\nonumber \\
=\frac{\pi^{-1}\Gamma(p+1)L_z^{2n}}{2^{n+3/2}\Gamma(n+\frac{1}{2})\Gamma(p-n-\frac{1}{2})}
\left(\varepsilon-\frac{L_z^2}{2}\right)^{p-n-3/2} \hspace*{1.5cm}
\label{ddf0s}
\end{eqnarray} 
for $\varepsilon>L_z^2/2$ and $f_+(\varepsilon,L_z)=0$ for $\varepsilon<L_z^2/2.$ 
It can be easily found that the DF (\ref{ddf0s}) is the same as obtained by use of (\ref{dfaq0}) when $p-n>1.$

Similar to those in {Sect.}~\ref{edf}, the velocity dispersions $\sigma_R^2(\psi,R)$ 
and $\sigma_\varphi^2(\psi,R)$ can be also found to be of the following forms 
\begin{equation}
\sigma_R^2(\psi,R)=\frac{1}{\rho(\psi,R)}\sum\limits_{n=0}^{m}
\frac{R^{2n}}{(1+R^2/R_a^2)^{n+1/2}}\int_0^\psi \hat{\rho}_n(\psi^\prime)d\psi^\prime
\label{vdraq}
\end{equation} 
and 
\begin{equation}
\sigma_\varphi^2(\psi,R)=\frac{1}{\rho(\psi,R)}
\sum\limits_{n=0}^{m}(2n+1)\frac{R^{2n}}{(1+R^2/R_a^2)^{n+3/2}}\int_0^\psi \hat{\rho}_n(\psi^\prime)d\psi^\prime
-\bar{v}_\varphi^2
\label{vdvtaq}
\end{equation} 
for  any DF derived from the axisymmetric density 
$\rho(\psi,R)$ defined by (\ref{rho2}).   

\subsection{Miscellaneous DFs}
\label{edfg} 
One can also obtain more general formulae than (\ref{dfa}) and (\ref{dfaq}). 
Assume that $\hat{Q}=\max(Q,0).$ Then 
it can be further shown that the DFs of the form 
\begin{eqnarray}
f_+(\varepsilon,Q,L_z)=&
\sum\limits_{n=0}^{m}B_{1n}L_z^{2n\beta_{1n}}
\left[\int_0^\varepsilon\frac{d^{a_{1n}+1}\tilde{\rho}_n(\psi)}
{d\psi^{a_{1n}+1}}\frac{d\psi}{(\varepsilon-\psi)^{\alpha_{1n}}}
+\frac{1}{\varepsilon^{\alpha_{1n}}}\left(\frac{d^{a_{1n}}\tilde{\rho}_n(\psi)}
{d\psi^{a_{1n}}}\right)_{\psi=0}\right] \hspace*{2.5cm} \nonumber \\
& + \sum\limits_{n=0}^{m}B_{2n}L_z^{2n\beta_{2n}}
\left[\int_0^{\hat{Q}}\frac{d^{a_{2n}+1}\hat{\rho}_n(\psi)}{d\psi^{a_{2n}+1}}\frac{d\psi}{(\hat{Q}-\psi)^{\alpha_{2n}}}
+\frac{1}{\hat{Q}^{\alpha_{2n}}}\left(\frac{d^{a_{2n}}\hat{\rho}_n(\psi)}{d\psi^{a_{2n}}}\right)_{\psi=0}\right]\hspace*{1cm}
\label{dfagqg}
\end{eqnarray}
correspond to an axisymmetric density of the form 
\begin{equation}
\rho(\psi,R)=\sum\limits_{n=0}^{m}
\tilde{\rho}_n(\psi)R^{2n\beta_{1n}}+ 
\sum\limits_{n=0}^{m}\hat{\rho}_n(\psi)R^{2n\beta_{2n}}/(1+R^2/R_a^2)^{n\beta_{2n}+1/2} 
\label{rhoagqg}
\end{equation}
 with $n\beta_{in}>-1,$ where
$B_{in}=[(2\pi)^{3/2}2^{n\beta_{in}}\Gamma(n\beta_{in}+1/2)\Gamma(1-\alpha_{in})]^{-1},$ 
$\alpha_{in}=n\beta_{in}-a_{in}+3/2$ and $a_{in}$ is 
a non-negative integer such that $0\leq\alpha_{in}<1$ for $i=1,2$ and $n=0,1,\cdots,m.$ 

Finally, the velocity dispersions $\sigma_R^2(\psi,R)$ and $\sigma_\varphi^2(\psi,R)$ can be also obtained as  
\begin{equation}
\sigma_R^2(\psi,R)=\frac{1}{\rho(\psi,R)}\sum\limits_{n=0}^{m}
\left[R^{2n\beta_{1n}}\int_0^\psi \tilde{\rho}_n(\psi^\prime)d\psi^\prime+
\frac{R^{2n\beta_{2n}}}{(1+R^2/R_a^2)^{n\beta_{2n}+1/2}}\int_0^\psi \hat{\rho}_n(\psi^\prime)d\psi^\prime\right] 
\label{vdraqg}
\end{equation} 
and 
\begin{equation}
\sigma_\varphi^2(\psi,R)=\frac{1}{\rho(\psi,R)}\sum\limits_{n=0}^{m}
\left[(2n\beta_{1n}+1)R^{2n\beta_{1n}}\int_0^\psi \tilde{\rho}_n(\psi^\prime)d\psi^\prime+
\frac{(2n\beta_{2n}+1)R^{2n\beta_{2n}}}{(1+R^2/R_a^2)^{n\beta_{2n}+3/2}}\int_0^\psi \hat{\rho}_n(\psi^\prime)d\psi^\prime\right]
-\bar{v}_\varphi^2
\label{vdvtaqg}
\end{equation} 
for  any DF derived from the axisymmetric density $\rho(\psi,R)$
given by  (\ref{rhoagqg}).  

\section{Models with gravitational potentials having no upper bound}
\label{mgpinf}
The axisymmetric gravitational potential $\Phi$ now has no upper bound and tends to $+\infty$ at large distances 
from which escape is impossible. Thus 
one usually denote by $f(E,L_z)$ the two-integral DF of a steady-state stellar system with the axisymmetric potential.  
Suppose that the system has only stars of $E>0.$ 
Then, by using the even two-integral DF $f_+(E,L_z)=[f(E,L_z)+f(E,-L_z)]/2,$ 
the fundamental integral equation (\ref{int0}) can be changed as follows:
\begin{equation}
 \rho=\frac{4\pi}{R}\int^{+\infty}_\Phi\left[\int_0^{R\sqrt{2(E-\Phi)}} 
f_+(E,L_z)dL_z\right]dE \label{intainf}.
\end{equation}
As in Section \ref{edf}, one can obtain the similar even DF
\begin{equation}
f_+(E,L_z)=\frac{1}{(2\pi)^{3/2}}
\sum\limits_{n=0}^{m}\frac{(-1)^nL_z^{2n}}{2^n\Gamma(n+\frac{1}{2})}\left[\frac{d}{dE}\int_E^{+\infty}
\frac{d^{n+1}\tilde{\rho}_n(\Phi)}{d\Phi^{n+1}}\frac{d\Phi}{\sqrt{\Phi-E}}\right]
\label{dfainf}
\end{equation}
for $E>0,$ 
corresponding to the axisymmetric density 
of the form $\rho(\Phi,R)=\sum\limits_{n=0}^{m}\tilde{\rho}_n(\Phi)R^{2n}$ under the assumption that 
\begin{equation}
  \lim\limits_{\Phi\to+\infty}\frac{d^j\tilde{\rho}_n(\Phi)}{d\Phi^j}=0 \hbox{  and  } 
 \lim\limits_{E\to+\infty}E\int_{E}^{+\infty}\frac{d^j\tilde{\rho}_n(\Phi)}{d\Phi^j}\frac{d\Phi}{\sqrt{\Phi-E}}=0 
\label{dfainfcon}
\end{equation}  
for all $j\in \{0,1,\dots,n\}$ and all $n\in \{0,1,2,\dots,m\}.$ 
Furthermore, if it is assumed that the condition (\ref{dfainfcon}) holds for all 
$j\in \{0,1,\dots,n+1\}$ and all $n\in \{0,1,2,\dots,m\},$ then, for $E>0,$ (\ref{dfainf}) can be expressed as 
\begin{equation}
f_+(E,L_z)=\frac{1}{(2\pi)^{3/2}}
\sum\limits_{n=0}^{m}\frac{(-1)^nL_z^{2n}}{2^n\Gamma(n+\frac{1}{2})}\int_E^{+\infty}
\frac{d^{n+2}\tilde{\rho}_n(\Phi)}{d\Phi^{n+2}}\frac{d\Phi}{\sqrt{\Phi-E}}.
\label{dfa0inf}
\end{equation}
It can be further shown that (\ref{dfainf}) and (\ref{dfa0inf}) are at least formally in accordance 
with the contour integrals given by Hunter and Qian (1993) for 
the gravitational potential tending to $+\infty$ at large distances. 

Put $Q=E+L_z^2/(2R_a^2)$ for the system with gravitational potentials having no upper bound. 
Similar to those given in Section \ref{edfq},
one can also obtain the even DF  
\begin{equation}
f_+(Q,L_z)=
\sum\limits_{n=0}^{m}\frac{(2\pi)^{-3/2}L_z^{2n}}{(-2)^n\Gamma(n+\frac{1}{2})}\frac{d}{dQ}\int_Q^{+\infty}
\frac{d^{n+1}\hat{\rho}_n(\Phi)}{d\Phi^{n+1}}\frac{d\Phi}{\sqrt{\Phi-Q}}
\label{dfaqinf}
\end{equation}
for $Q>0,$   corresponding to the axisymmetric density $\rho(\Phi,R)$ of the form defined by 
\begin{equation}
\rho(\Phi,R)=\sum\limits_{n=0}^{m}\hat{\rho}_n(\Phi)R^{2n}/(1+R^2/R_a^2)^{n+1/2}
\label{rho2q}
\end{equation}  
under the assumption that 
\begin{equation}
  \lim\limits_{\Phi\to+\infty}\frac{d^j\hat{\rho}_n(\Phi)}{d\Phi^j}=0 \hbox{  and  } 
 \lim\limits_{Q\to+\infty}Q\int_{Q}^{+\infty}\frac{d^j\hat{\rho}_n(\Phi)}{d\Phi^j}\frac{d\Phi}{\sqrt{\Phi-Q}}=0 
\label{dfainfconq}
\end{equation}  
hold for all 
$j\in \{0,1,\dots,n\}$ and all $n\in \{0,1,2,\dots,m\}.$
Furthermore, if one assume that the condition (\ref{dfainfconq}) holds for all 
$j\in \{0,1,\dots,n+1\}$ and all $n\in \{0,1,2,\dots,m\},$
 then for $Q>0,$ (\ref{dfaqinf}) can be rewritten as 
\begin{equation}
f_+(Q,L_z)=
\sum\limits_{n=0}^{m}\frac{(2\pi)^{-3/2}L_z^{2n}}{(-2)^n\Gamma(n+\frac{1}{2})}\int_Q^{+\infty}
\frac{d^{n+2}\hat{\rho}_n(\Phi)}{d\Phi^{n+2}}\frac{d\Phi}{\sqrt{\Phi-Q}}.
\label{dfaq0inf}
\end{equation}

As in Section \ref{edfg}, one can further show the more general DFs of the form 
\begin{eqnarray}
f_+(E,Q,L_z)= &
\sum\limits_{n=0}^{m}(-1)^{a_{1n}-1}B_{1n}L_z^{2n\beta_{1n}}
\frac{d}{dE}\int_E^{+\infty}\frac{d^{a_{1n}}\tilde{\rho}_n(\Phi)}
{d\Phi^{a_{1n}}}\frac{d\Phi}{(\Phi-E)^{\alpha_{1n}}} \nonumber\\ &
+ \sum\limits_{n=0}^{m}(-1)^{a_{2n}-1}B_{2n}L_z^{2n\beta_{2n}}
\frac{d}{d\hat{Q}}\int_{\hat{Q}}^{+\infty}\frac{d^{a_{2n}}\hat{\rho}_n(\Phi)}{d\Phi^{a_{2n}}}\frac{d\Phi}{(\Phi-\hat{Q})^{\alpha_{2n}}}
\label{dfagqginf}
\end{eqnarray}
corresponding to an axisymmetric density of the form given by 
\begin{equation}
\rho(\Phi,R)=\sum\limits_{n=0}^{m}
\tilde{\rho}_n(\Phi)R^{2n\beta_{1n}}+ 
\sum\limits_{n=0}^{m}\hat{\rho}_n(\Phi)R^{2n\beta_{2n}}/(1+R^2/R_a^2)^{n\beta_{2n}+1/2}, 
\label{rhoagqginf}
\end{equation}
where  $\hat{Q}=\max(Q,0),$ 
$a_{in},$ $B_{in},$ $\alpha_{in}$ and $\beta_{in}$ are the same as in (\ref{dfagqg}) for $i=1,2.$   

By (\ref{intainf}), it can be also found that, in the system with only stars of $E>0,$ the even DF 
\begin{equation}
f_+(E,L_z)=L_z^{2n+1}\exp\left(-\alpha E-\frac{\beta L_z^2}{2R_0^2}\right)
\label{dfaexpinf}
\end{equation}
corresponds to an axisymmetric density of the form 
\begin{equation}
\rho(\Phi,R)=\frac{4\pi(2n)!!R_0^{2(n+1)}R^{2n+1}e^{-\alpha \Phi}}
{\alpha (R_0^2\alpha+\beta R^2)^{n+1}}
\label{rhoexpinf}
\end{equation}
for any gravitational potential tending to $+\infty,$ where $\alpha$ and $\beta$ are nonnegative constants, $R_0$ is a positive constant, 
$n$ is a nonnegative integer,  
and $(2n)!!=1\cdot 2\cdot 4\cdots (2n-2)\cdot (2n)$ when $n$ is a natural number 
and $(0)!!$ is defined to be equal to one. 
It is very remarkable that (\ref{dfaexpinf}) can be recovered 
from the complex contour integral given by Hunter and Qian (1993).  
It can be below used to find the odd part of the DF that 
corresponds to some assumed rotational velocity $\langle v_\phi\rangle$ 
for the axisymmetric stellar systems. 

\section{Application to the axisymmetric cases}
\label{aac}
Binney's (BT) logarithmic model has infinite mass and its gravitational potential is of the form
\begin{equation}
\Phi(R,z)=\frac{1}{2}v_0^2\ln\left(1+R^2+\frac{z^2}{q^2}\right),
\label{psiab}
\end{equation}
where $v_0$ is the constant circular velocity in the equatorial plane at large distances, 
$q$ is the axial ratio of the spheroidal equipotentials. 
Obviously, this gravitational potential has no upper bound. 
 The density derived from (\ref{psiab}) is 
\begin{equation}
\rho(R,z)=\frac{v_0^2}{4\pi Gq^2}\left\{2[(1-q^2)R^2+1]e^{-4\Phi/v_0^2}+(2q^2-1)e^{-2\Phi/v_0^2}\right\}.
\label{rhoab}
\end{equation}
Then, by (\ref{dfa0inf}), one can obtain the even DF corresponding to (\ref{rhoab}) as follows:
\begin{equation}
f_+(E,L_z)=\frac{1}{4\pi Gq^2v_0^3}
\left\{2^{9/2}[(1-q^2)L_z^2+2^{5/2}v_0^2]e^{-4E/v_0^2}+(2q^2-1)v_0^2e^{-2E/v_0^2}\right\},
\label{dfab}
\end{equation}
which is the same as found by Evans (1993) using Lynden-Bell's (1962) method. 
This kind of solution was known earlier to Toomre (1982) and published first by Miller (1982). 
It can be easily found that the mass density (\ref{rhoab}) of the Binney model is positive 
in the position space only 
when the axial ratio $q$ is not less than $1/\sqrt{2}.$  
It is easy to see that,  
in the prolate case when the axial ratio $q$ is greater than $1,$  
the DFs (\ref{dfab}) must be negative at some points in the physical domain 
and so they are not the real two-integral DFs for the stellar system. 
When the axial ratio $q$ is equal to one,  the model is spherical. 
The model is flattened only when $1>q\geq1/\sqrt{2}\approx 0.707107.$
Figure~\ref{bmplot} illustrates the contours of the DFs given by (\ref{dfab}) with three different axial ratios. One of them is 
for the spherical case and the other two for the flattened one.   
\setcounter{figure}{0}
\begin{figure}
%\vbox to110mm{\vfil 
\centering\includegraphics{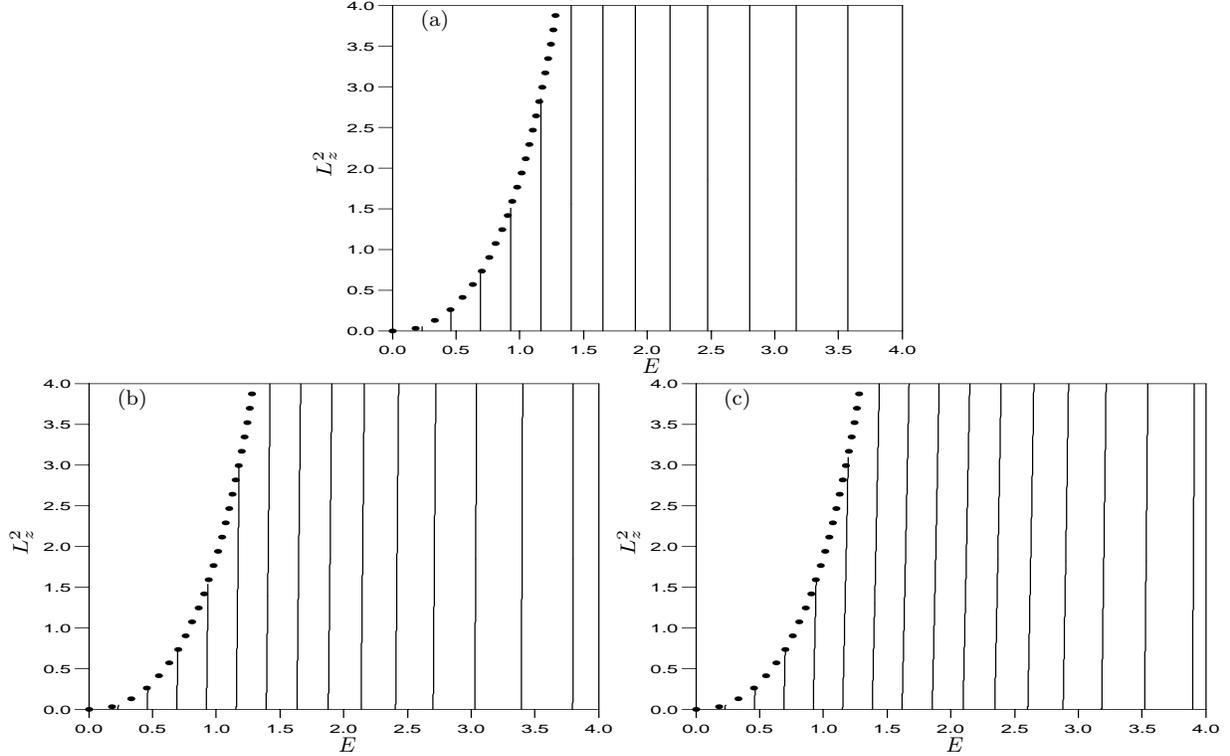}
%\psfrag{E}{\small $E$}\psfrag{Lz2}{\small $L_z^2$}
%\psfrag{a}{\small (a)}\psfrag{b}{\small (b)}\psfrag{c}{\small (c)}
%\centering
%\includegraphics[bb=16 16 211 222,totalheight=50mm,width=0.45\linewidth]{sphericalc.eps}\\
%\includegraphics[bb=16 16 211 222,totalheight=50mm,width=0.45\linewidth]{qp9c.eps}
%\includegraphics[bb=16 16 211 222,totalheight=50mm,width=0.45\linewidth]{qp8c.eps}
\caption{The contours of the DFs given by (\ref{dfab}) with three different axial ratios when $v_0$ is set to be one.  
Subfigures (a), (b) and (c) are for $q=1,$ $q=0.9$ and $q=0.8,$ respectively. 
The thin solid curves are isocontours and the dotted curve is the boundary of the physical domain. 
Successive levels differ by factors of $0.4.$}
%\vfil}
\label{bmplot}
\end{figure}

Note that if $\rho$ and $f_+(E,L_z)$ in (\ref{intainf}) 
are replaced by $\rho R \langle v_\phi\rangle$ and $L_zf_\_(E,L_z),$ respectively, 
then (\ref{intainf}) becomes an equation for $L_zf_\_(E,L_z),$ that is, 
\begin{equation}
 \rho R \langle v_\phi\rangle=\frac{4\pi}{R}\int^{+\infty}_\Phi\left[\int_0^{R\sqrt{2(E-\Phi)}} 
L_zf_\_(E,L_z)dL_z\right]dE \label{intaoddinf},
\end{equation}
where $f_\_(E,L_z)$ is usually called the odd DF  
 given by $f_\_(E,L_z)=[f(E,L_z)-f(E,-L_z)]/2$  
 for the stellar systems.  
This property was first found by Lynden-Bell (1962) and then applied by Hunter and Qian (1993) into 
calculating the odd part of the DF for the Binney model under the assumption of 
the rotational velocity $\langle v_\phi\rangle$ having the rotation law: 
\begin{equation}
\langle v_\phi\rangle=v_\ast R^2/(R_\ast^2+R^2),
\label{rot-law}
\end{equation} 
where $v_\ast$ and $R_\ast$ are constant velocity and length scales, respectively. 
(\ref{rot-law}) was one of the rotation laws considered by Evans (1993). 
Using the Hunter and Qian's (1993) contour integral formulae of the even DF, Hunter \& Qian (1993) 
first derived a contour integral of the 
odd DF from the rotational velocity $\langle v_\phi\rangle$ for the Binney model and this idea 
was then extended by Jiang (2000) 
to the odd DF for a more general model. 
By combining (\ref{dfaexpinf}) and (\ref{intaoddinf}), it can be also found that, in the system with only stars of $E>0,$ the odd DF 
\begin{equation}
f_\_(E,L_z)={\rm sgn}(L_z)L_z^{2n}\exp\left(-\alpha E-\frac{\beta L_z^2}{2R_0^2}\right)
\label{dfaexpoddinf}
\end{equation}
corresponds to an axisymmetric density $\rho(\Phi,R)$ of the form 
\begin{equation}
\rho(\Phi,R)\langle v_\phi\rangle=\frac{4\pi(2n)!!R_0^{2(n+1)}R^{2n}e^{-\alpha \Phi}}
{\alpha (R_0^2\alpha+\beta R^2)^{n+1}}
\label{rhoexpoddinf}
\end{equation}
for any system with a gravitational potential tending to $+\infty,$ where $\alpha,$ $\beta,$ $R_0,$  
$n$ and $(2n)!!$ are the same as in (\ref{dfaexpinf}). 
If one assumes that the rotational velocity $\langle v_\phi\rangle$ of 
the Binney model satisfies the rotation law 
\begin{equation}
\langle v_\phi\rangle=v_\ast R^{2(n+1)}/(R_\ast^2+R^2)^{n+1}
\label{rot-lawg}
\end{equation}
for any nonnegative integer $n,$ where $v_\ast$ and $R_\ast$ are the same as in (\ref{rot-law}), 
then, combining (\ref{rhoab}), (\ref{dfaexpoddinf}) and (\ref{rhoexpoddinf}), one can find that the odd DF is given by 
\begin{eqnarray}
f_\_(E,L_z)=\frac{v_\ast {\rm sgn}(L_z)}{4\pi^2Gq^2v_0^2}
\left\{16(1-q^2)\frac{L_z^2}{v_0^2}\exp\left(-\frac{4E}{v_0^2}\right)
+8[1-(n+1)R_\ast^2(1-q^2)]\exp\left(-\frac{4E}{v_0^2}\right)\right. \nonumber \\ 
-8\sum\limits_{j=0}^{n}\frac{2^{2j}}{(2j)!!}\left(\frac{L_z}{R_\ast v_0}\right)^{2j}\exp\left(-\frac{4E}{v_0^2}-\frac{2L_z^2}{R_\ast^2v_0^2}\right)
+(2q^2-1)\exp\left(-\frac{2E}{v_0^2}\right)  \nonumber \\
+16R_\ast^2(1-q^2){\rm sgn}(n)\sum\limits_{k=1}^{n/2}\sum\limits_{j=0}^{2k-1}
\frac{2^{2j}}{(2j)!!}\left(\frac{L_z}{R_\ast v_0}\right)^{2j}\exp\left(-\frac{4E}{v_0^2}-\frac{2L_z^2}{R_\ast^2v_0^2}\right)  \nonumber \\
+8R_\ast^2(1-q^2)\sum\limits_{j=0}^{n/2}
\frac{2^{4j}}{(4j)!!}\left(\frac{L_z}{R_\ast v_0}\right)^{4j}\exp\left(-\frac{4E}{v_0^2}-\frac{2L_z^2}{R_\ast^2v_0^2}\right)  \nonumber \\
\left.+(1-2q^2)\sum\limits_{j=0}^{n}\frac{2^{j}}{(2j)!!}\left(\frac{L_z}{R_\ast v_0}\right)^{2j}\exp\left(-\frac{2E}{v_0^2}-\frac{L_z^2}{R_\ast^2v_0^2}\right)
\right\}
\label{odddfe}
\end{eqnarray}
for any nonnegative even number $n$ and that the odd DF is expressed as 
\begin{eqnarray}
f_\_(E,L_z)=\frac{v_\ast {\rm sgn}(L_z)}{4\pi^2Gq^2v_0^2}
\left\{16(1-q^2)\frac{L_z^2}{v_0^2}\exp\left(-\frac{4E}{v_0^2}\right)
+8[1-(n+1)R_\ast^2(1-q^2)]\exp\left(-\frac{4E}{v_0^2}\right)\right. \nonumber \\ 
-8\sum\limits_{j=0}^{n}\frac{2^{2j}}{(2j)!!}\left(\frac{L_z}{R_\ast v_0}\right)^{2j}\exp\left(-\frac{4E}{v_0^2}-\frac{2L_z^2}{R_\ast^2v_0^2}\right)
+(2q^2-1)\exp\left(-\frac{2E}{v_0^2}\right)  \nonumber \\
+16R_\ast^2(1-q^2)\sum\limits_{k=0}^{(n-1)/2}\sum\limits_{j=0}^{2k}
\frac{2^{2j}}{(2j)!!}\left(\frac{L_z}{R_\ast v_0}\right)^{2j}\exp\left(-\frac{4E}{v_0^2}-\frac{2L_z^2}{R_\ast^2v_0^2}\right)  \nonumber \\
+8R_\ast^2(1-q^2)\sum\limits_{j=0}^{(n-1)/2}
\frac{2^{4j+2}}{(4j+2)!!}\left(\frac{L_z}{R_\ast v_0}\right)^{4j+2}\exp\left(-\frac{4E}{v_0^2}-\frac{2L_z^2}{R_\ast^2v_0^2}\right)  \nonumber \\
\left.+(1-2q^2)\sum\limits_{j=0}^{n}\frac{2^{j}}{(2j)!!}\left(\frac{L_z}{R_\ast v_0}\right)^{2j}\exp\left(-\frac{2E}{v_0^2}-\frac{L_z^2}{R_\ast^2v_0^2}\right)
\right\}
\label{odddfo}
\end{eqnarray}
for any positive odd number $n.$
This also means that there are an infinity of the DFs for any given axisymmetric stellar potential. 
(\ref{rot-lawg}) is obviously an extension of (\ref{rot-law}), and when $n=0,$ 
(\ref{odddfe}) is the same as given by Hunter and Qian (1993). 

The well-known Lynden-Bell (1962) model has finite mass and its relative potential and density are given by 
\begin{equation}
\psi(R,z)=[(R^2+z^2+1)^2+aR^2]^{-1/4},
\label{psialb}
\end{equation}
\begin{equation}
\rho(R,z)=\frac{\psi^5}{4\pi G}\left[(3+a)-5a\left(1+\frac{a}{4}\right)R^2\psi^4\right],
\label{rhoalb}
\end{equation}
where $a$ is a flattening parameter. By using (\ref{dfa}) and (\ref{rhoalb}), 
it follows that the even DF is given by 
\begin{equation}
f_+(\varepsilon,L_z)=\frac{1}{2^{3/2}\pi^2}\varepsilon^{7/2}
\left(-\frac{15a(4+a)2^{12}}{143}\varepsilon^3L_z^2+\frac{2^7(3+a)}{7}\right),
\label{dfalb}
\end{equation}
which is in fact as the method of Fricke (1952). 
It is here necessary to explain the different definitions of the gravitational potential. 
In the paper written by Lynden-Bell (1962), (\ref{psialb}) is called the gravitational 
potential of the Lynden-Bell model.  Due to the use of the concept of the relative potential, 
the gravitational potential defined by Binney and Tremaine (BT)  in fact differs by a factor $-1$ from 
that given by Lynden-Bell (1962). The Binney and Tremaine definition of 
the gravitational potential is used throughout this paper and so (\ref{psialb}) is a relative potential. 

\section{Conclusions}
\label{con}
Few galaxies are even nearly spherical.  
Thus it is a natural idea to 
explore some important properties of real galaxies  
 by employing the cylindrical polar coordinate system $(R,\varphi,z)$ 
with the center on the galactic nucleus 
and the $z$-axis being that of symmetry of the galaxy. 
However, at least two involved factors require being mentioned as follows.  
One is that some possible orbits in many real galaxies can be easily described by studying  
a two-dimensional problem. 
 With the help of the conservation of the angular momentum about the symmetry $z$-axis, 
this problem can be directly reduced from the analysis of the orbits 
in the three-dimensional space occupied by axisymmetric galaxies.  
The other is that, on the analogy of anisotropic DFs for spherical systems, 
the two-integral DFs for some axisymmetric systems can be also found to 
model the typical behaviours of the dynamical quantities of galaxies considered. 

Some formulae of the two-integral DFs can be obtained for  stellar systems
with known axisymmetric density as a sum of products 
of functions only of the potential and 
a  special function (or power) only of the radial coordinate, 
i.e. these DFs are a sum of products of functions 
only of a special variable (or the energy)  
and a power only of the 
magnitude of the angular momentum about the axis of symmetry.  
They come from an combination of the ideas of 
 Eddington and Fricke and they 
are also an extension of those shown by Jiang and Ossipkov (2007) for  
finding anisotropic distribution functions for spherical galaxies.  
As an analogue for spherical models, 
the product of the density and its radial velocity dispersion   
can be also expressed as  
  a sum of products of functions of the potential and 
 of the radial coordinate. But the expression of its rotational velocity dispersion formally 
differs from that of its radial velocity dispersion. It can be further found that   
the density multiplied by the difference between 
 the dispersion of its rotational velocity and the square of its mean rotational velocity 
is equal to a sum of products of functions of the potential and 
 of the radial coordinate. 
The similar formulae of the two-integral DFs 
for the gravitational potentials without upper bound 
 are as well in accordance with the complex contour integral ones given by Hunter \& Qian (1993). 
These expressions for axisymmetric systems  
can be used to obtain the even DF of Binney's (BT) logarithmic 
potential although Evans (1993) derived it using Lynden-Bell's (1962) method. 
An infinity of the odd DFs for the Binney model can be also found under the assumption of 
the laws of the rotational velocity. 
For the well-known Lynden-Bell (1962) model,  
these analogues degenerate into the method of Fricke (1952). 
It is worth mentioning that such analytic procedure to determine the DFs 
can be also applied to the prolate Jaffe models given by Jiang \& Moss (2002) for 
 a good numerical approximation of the two-integral DFs for the stellar systems.

One can finally know that it is a shortcoming of  all the two-integral models that the radial velocity dispersion  
is equal to the vertical velocity dispersion. This is because  
it is well-known that in real axisymmetric stellar systems, the velocity dispersion in the radial direction 
is not equal to the velocity dispersion in the vertical direction, meaning that the DFs of the real systems 
must actually depend on three integrals of the motion (one of them being non-analytic in general) 
rather than two (of course, there is then no unique solution for the even part of the DF). 
To  overcome this shortcoming of the two-integral models, some extensions of two-integral DFs 
have been studied to construct three-integral DFs for particular orbital families in flattened axisymmetric systems 
(Evans, Hafner \& de Zeeuw 1997) and for separable axisymmetric St\"{a}ckel potentials (Famaey, Van Caelenberg \& Dejonghe 2002). 

%\begin{acknowledgement}
\vskip0.2cm
\noindent {\bf Acknowledgement}.  
The first author was supported by NSFC 10271121 and by 
SRF for ROCS, SEM. 
The first author would like to thank Dr.~David Moss for his 
helpful comments on this paper.  
The second author was supported by Leading Scientific School grant 1078.2003.02. 
The cooperation of authors was supported by joint grants 
of NSFC 10511120278/10611120371 and RFBR 04-02-39026. 
The two authors are very grateful to Professor Konstantin Kholshevnikov, Professor Sergei Kutuzov 
and Professor Vadim Antonov for their valuable discussions on this work.  
The two authors would also like to thank the referee of this paper 
for his/her valuable comments on this work.
%\end{acknowledgement}

\bsp

\label{lastpage}

\end{document}